\documentclass{ieeeaccess-fixed}
\usepackage{draftwatermark}
\SetWatermarkText{Accepted Manuscript}
\SetWatermarkScale{0.5}

\usepackage{cite}
\usepackage{amsmath,amssymb,amsfonts}
\usepackage{algorithmic}
\usepackage{graphicx}
\usepackage{textcomp}

\usepackage{bm}
\makeatletter
\AtBeginDocument{\DeclareMathVersion{bold}
\SetSymbolFont{operators}{bold}{T1}{times}{b}{n}
\SetSymbolFont{NewLetters}{bold}{T1}{times}{b}{it}
\SetMathAlphabet{\mathrm}{bold}{T1}{times}{b}{n}
\SetMathAlphabet{\mathit}{bold}{T1}{times}{b}{it}
\SetMathAlphabet{\mathbf}{bold}{T1}{times}{b}{n}
\SetMathAlphabet{\mathtt}{bold}{OT1}{pcr}{b}{n}
\SetSymbolFont{symbols}{bold}{OMS}{cmsy}{b}{n}
\renewcommand\boldmath{\@nomath\boldmath\mathversion{bold}}}
\makeatother

\def\BibTeX{{\rm B\kern-.05em{\sc i\kern-.025em b}\kern-.08em
    T\kern-.1667em\lower.7ex\hbox{E}\kern-.125emX}}

\usepackage[]{CJK}

\usepackage{hyperref}
\hypersetup{
    colorlinks=true,
    citecolor=blue,
    linkcolor=blue,
    filecolor=magenta,
    urlcolor=blue,
    }

\usepackage{ragged2e}

\begin{document}
\history{Date of publication: June 4th, 2026}
\doi{\href{https://doi.org/10.48550/arXiv.2512.07353}{10.48550/arXiv.2512.07353}}

\title{Off-grid solar energy storage system with hybrid lithium iron phosphate (LFP) and lead-acid batteries in high mountains: a case report of Jiujiu Cabins in Taiwan}
\author{\uppercase{Hsien-Ching Chung}\authorrefmark{\href{https://orcid.org/0000-0001-9364-8858}
{\includegraphics[height=6.5pt]{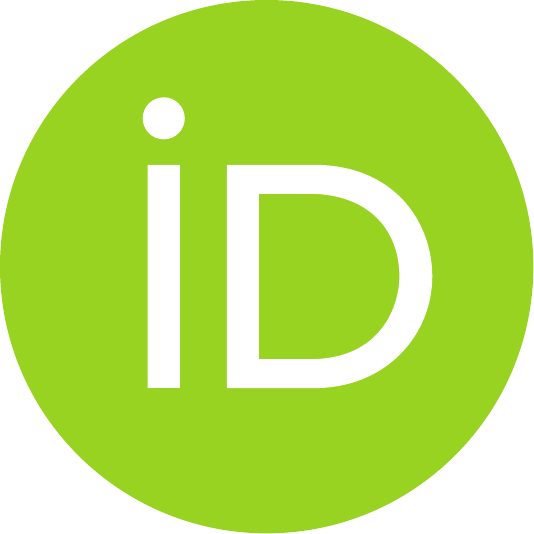}}1},
\IEEEmembership{Senior Member, IEEE}
}

\address[1]{Department of Research and Design, Super Double Power Technology Co., Ltd., Changhua City, Changhua County 500042, Taiwan (e-mail: hsienching.chung@gmail.com)}


\corresp{Corresponding author: Hsien-Ching Chung (e-mail: hsienching.chung@gmail.com).}

\begin{abstract}
Mountain huts are buildings located at high altitude, offering a place for hikers and providing shelter. Energy supply on mountain huts is still an open issue. Using renewable energies could be an appropriate solution. Jiujiu Cabins, a famous mountain hut in Shei-Pa National Park, Taiwan, has operated an off-grid solar energy storage system (ESS) with lead-acid batteries. In 2021, a serious system failure took place, leading to no electricity. After a detailed on-site survey, a reorganization and repair project implemented, the energy system came back to operate normally. Meanwhile, an eco-friendly lithium iron phosphate battery (LFP battery) ESS replaces part of the lead-acid battery ESS, forming a hybrid ESS, making a better and green off-grid solar ESS. In this case report, the energy architecture, detailed descriptions, and historical status of the system are provided. An on-site survey of the failed energy system, a system improvement project, and future plan are listed.
\end{abstract}


\titlepgskip=-21pt

\maketitle

\section{Introduction}
\label{sec:introduction}

Mountain huts are buildings located at high altitude, in mountainous terrain, generally accessible only by foot, intended to provide food and shelter to mountaineers, hikers, and climbers~\cite{CanadianGeographies36(1992)144H.G.Kariel}.
They are known by many names, such as alpine hut, mountain shelter, mountain refuge, mountain lodge, mountain hostel, and mountain cabin.
Mountain huts can provide some functionalities and services.
The most important function is to provide shelter and simple sleeping berths for hikers to overcome the cold mountain environment at night.
Many mountain huts are not staffed, particularly in remote areas. However, others have staff who provide hot meals, drinks, and other services, such as activity spaces (for holding lectures, etc.), clothing sales, and small items.

Energy supply in high mountains remains an open issue to be solved.
Generally, grid connection is impractical, because the establishment and maintenance costs are too high.
It also causes many environmental problems, e.g., the destruction of the natural ecology in the mountain.
Diesel generators as energy suppliers have been used in high mountains for a long time despite their air pollution and noise problems.
Due to the development of renewable energy (such as solar, wind, and hydropower), the usage of diesel generators is reduced, lowering the emissions of greenhouse gases (GHGs).
However, owing to their fluctuating nature, most renewable energy sources exhibit intermittent features.
To deal with the problem of renewable energy intermittency, energy storage systems (ESSs) are necessary~\cite{IEEETrans.SmartGrid6(2015)124K.Rahbar, IEEETrans.Sustain.Energy1(2010)117S.Teleke}.
In the past, lead-acid batteries were heavily used as ESSs, accompanied by many environmental issues, e.g., poisoning, Pb leaks, contamination of the environment, and damage to the ecosystem~\cite{Ecol.Indic.47(2014)210G.N.Liu, J.Hazard.Mater.250-251(2013)387X.F.Zhu}.
Recently, lithium-ion (Li-ion) batteries~\cite{EnergyStorageMater.1(2015)158J.B.Goodenough, Chem.Mater.22(2010)587J.B.Goodenough}, as a greener alternative, have started to replace lead-acid batteries in ESSs.
Lithium iron phosphate (LFP) batteries have a slightly lower energy density compared to other Li-ion cell chemistries due to their lower operating voltage. Their special features, such as low cost, low toxicity, low self-discharge, high cycle life~\cite{Sustainability11(2019)2527C.S.Ioakimidis, Nanomaterials13(2023)1486P.Cofre, Batteries10(2024)137W.Wheeler}, high power, and high thermal stability, make them find many roles in vehicle usage~\cite{Transp.Res.Proc.33(2018)195I.Carrilero} and utility-scale stationary application~\cite{Energies9(2016)887F.M.Gatta, IEEETrans.Sustain.Energy8(2017)385B.Lian}.

In order to ensure energy availability in high-altitude regions, some research is attempting to introduce new technologies.
Cabezas et al. demonstrated that hydrogen energy storage (HES) can be used in low-pressure areas, indicating that HES could be used in high mountains~\cite{Int.J.HydrogenEnergy39(2014)18165M.D.Cabezas}.
Chung demonstrated that the LFP ESS installed in Paiyun Lodge on Mt. Jade, Taiwan (elevation 3,402 m) can operate for more than 7 years~\cite{Batteries10(2024)202H.C.Chung}.
Alberizzi et al. developed an optimization model for using hybrid renewable energy systems where several energy sources can be incorporated.
The developed model has been applied to a case study of a mountain hut (elevation 2,200 m), located in the Italian region of South-Tyrol (Italy), in order to assess the optimal sizing of a photovoltaic-wind-diesel generator system, coupled with a lead-acid battery storage~\cite{EnergyConvers.Manag.223(2020)113303J.C.Alberizzi}.
Mori et al. have studied several configurations of mountain-hut stand-alone energy systems and pointed out the advantages of using HES in high mountains~\cite{Int.J.HydrogenEnergy46(2021)29706M.Mori, Energies15(2022)202M.Mori}.

Jiujiu Cabins, a famous mountain hut located in Shei-Pa National Park, Taiwan, has adopted an off-grid solar ESS to provide electricity.
However, lead-acid batteries were implemented as an ESS, accompanied by many environmental issues.
In 2021, owing to the aging of lead-acid batteries and equipment failures, electricity was unavailable for a short period.
After a detailed on-site survey, a solar power system improvement project was initiated.
The repaired energy system came back to provide electricity normally.
Moreover, the lead-acid batteries were partly replaced by eco-friendly LFP batteries, forming a hybrid ESS composed of lead-acid batteries and LFP batteries.
A better and greener off-grid solar energy storage system has been established.

In this manuscript, a brief introduction of the Jiujiu Cabins is given in Sec.~\ref{sec:AboutJiujiuCabins}.
In Sec.~\ref{sec:ArchitectureEnergySystem}, the current status of the energy architecture of the off-grid solar ESS is presented with a detailed description of each component.
A historical development of energy systems is listed in Sec.~\ref{sec:HistoricalDevelopmentEnergySystems}.
In 2021, Jiujiu Cabins experienced a brief period of power outage, and the investigation results of the system failure are listed in Sec.~\ref{sec:On-siteSurveyEnergySystems}.
How to overhaul and improve the system, and the engineering work to restore power, is covered in Sec.~\ref{sec:SolarPowerSystemImprovementProject2021}.
Possible future system upgrade plans are listed in Sec.~\ref{sec:FuturePlanningForEnergySystem}.
Summary and Outlook are given at last (Sec.~\ref{sec:SummaryOutlook}).


\section{About Jiujiu Cabins}
\label{sec:AboutJiujiuCabins}

\subsection{Current status}

\begin{figure}[htb]
  \centering
  \includegraphics[width=\columnwidth]{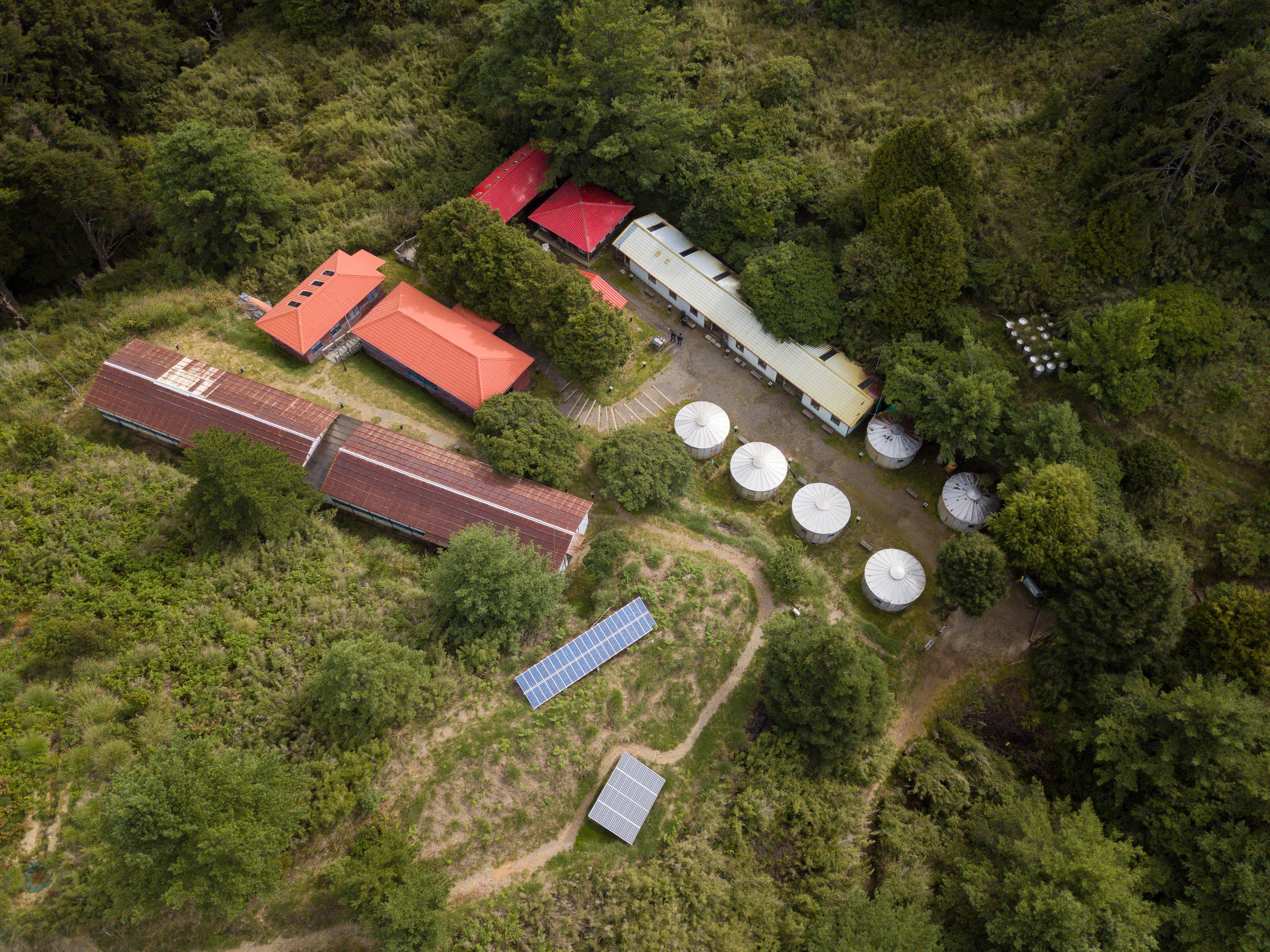}
  \caption{
  \justifying
  \textbf{Aerial photography of Jiujiu Cabins (2021).}
  The cabin complex.
  It includes 6 Chengkung Yurts, a restaurant (reception room), 3 Longmen Inns, a machine room, a toilet, etc., all of which are separate buildings.
  (Photographer: Yu-Chun Lin)
  }
  \label{fig:JiujiuCabinsAerialPhotography}
\end{figure}

Jiujiu Cabins (coordinate: $24^\circ 28^\prime 08.7^{\prime\prime}$N, $121^\circ 12^\prime 29.9^{\prime\prime}$E) is an accommodation mountain hut in Taiwan, located in Shei-Pa National Park, in Tai'an Township, Miaoli County, at an elevation of 2,699 m (8,855 ft).
It is named after the number 99 in its elevation~\cite{JiujiuCabinsAboutUs()TW-HsinchuFNCA}.
The cabin complex belongs to the Hsinchu Branch, Forestry and Nature Conservation Agency, Ministry of Agriculture.
It includes 6 Chengkung Yurts, a restaurant (reception room), 3 Longmen Inns, a machine room, a toilet, etc., all of which are separate buildings (as shown in Fig.~\ref{fig:JiujiuCabinsAerialPhotography}).
Jiujiu Cabins uses a photovoltaic system and is equipped with a Li-ion battery ESS to cope with the intermittent nature of renewable energy~\cite{JiujiuCabinsAboutUs()TW-HsinchuFNCA}.
It can currently accommodate 150 people~\cite{JiujiuCabins150Person(2019)TW-Shei-PaHead, JiujiuCabins150Person(2019)TW-LTN}.
It has clean running water (but currently no hot water provided) and a mobile phone communication point~\cite{Shei-Pa-CellPhoneSignal-90Areas(2022)TW-CNA}.

\begin{figure}[htb]
  \centering
  \includegraphics[width=\columnwidth]{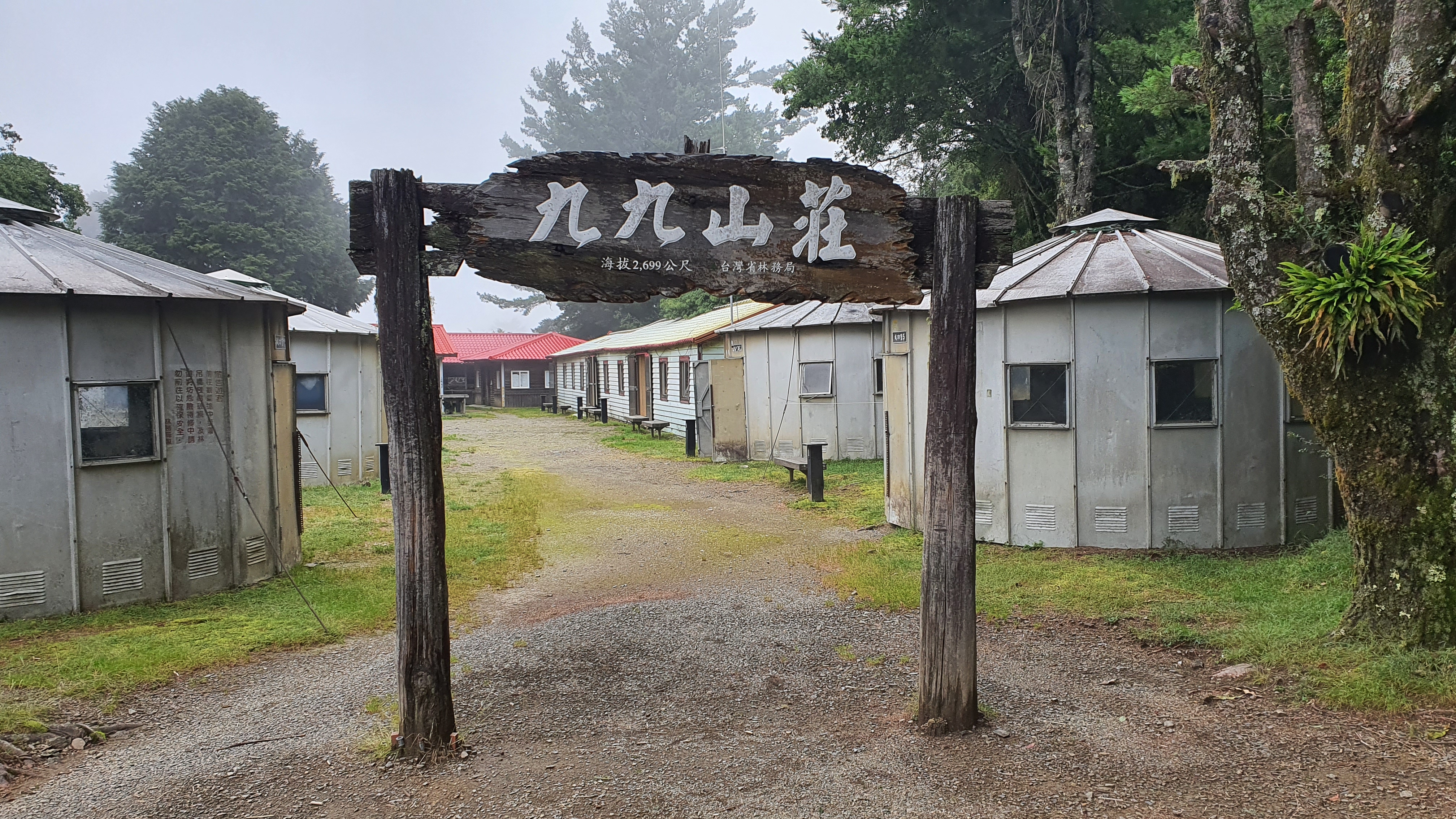}
  \caption{
  \justifying
  \textbf{Entrance of Jiujiu Cabins (2021).}
  }
  \label{fig:EntranceJiujiuCabins}
\end{figure}

\subsection{Brief history}

In the 1960s, Jiujiu Cabins was built with lead sheets in the shape of a yurt, which is a common memory of mountain climbers~\cite{JiujiuCabinsPhoto(1960)TW-MOC} and was later called ``Chengkung Yurt.''
These six Chengkung Yurts are made of metal and are round mountain houses in the shape of yurts.

Since 1971, in order to cooperate with the China Youth Corps in organizing the Dabajian Mountain hiking event, the Forestry Bureau (now the Forestry and Nature Conservation Agency, Ministry of Agriculture, Taiwan) has opened a hiking trail about 12 kilometers long from the Dalu Forest Road Madala Creek Trailhead to Dabajian Mountain. In addition, Jiujiu Cabins were built in the mountain valley on the west side of the shoulder ridge in the northwest of Jiali Mountain to provide accommodation for the hiking team~\cite{MountSylvaniaStory()CTBC-Museum, MountSylvaniaStory()TW-Sunriver, MountSylvaniaStoryWeb()TW-Sunriver}. To date, there are 6 Chengkung Yurts and 3 Longmen Inns as accommodation spaces, all of which are dormitory-style (with soft mattresses). Hikers need to bring their own sleeping bags and other equipment~\cite{JiujiuCabinsAboutUs()TW-HsinchuFNCA}.

\subsection{Geography}

Jiujiu Cabins are an important place for mountain enthusiasts to stay and rest on their way to Daba Peaks (Dabajian Mountain, Xiaobajian Mountain, Yizhe Mountain, Jiali Mountain, etc.)~\cite{JiujiuCabinsEatingService(2021)TW-LTN}. They usually drive from Dalu Forest Road to Guanwu National Forest Recreation Area~\cite{JiujiuCabinsPhoto(1960)TW-MOC}, and then hike along the flat forest road to the 19 km mark of Madala Creek Trailhead, and then continue for another 4 km along the mountain trail to reach Jiujiu Cabins~\cite{IntroductionDabaPeaksTrail()JiujiuCabins}.

\section{Architecture of current energy system}
\label{sec:ArchitectureEnergySystem}

Due to the difficulty of power grid connection in high-altitude areas, energy supply remains a challenging problem. Off-grid solar ESSs~\cite{Batteries10(2024)202H.C.Chung, BookChChung2021EngIntePotentialAppOutlooksLiIonBatteryIndustry} provide an appropriate solution for using renewable energy in high-altitude areas and have been proven~\cite{Batteries10(2024)202H.C.Chung}.

\begin{figure}[ht]
  \centering
  \includegraphics[width=\columnwidth]{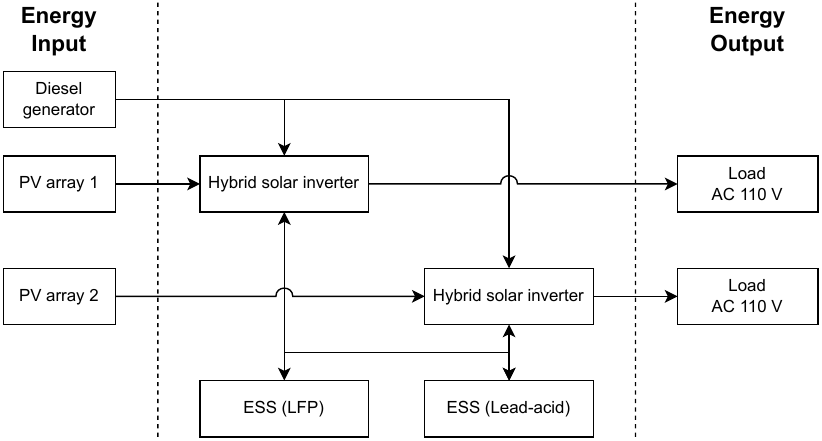}
  \caption{
  \justifying
  \textbf{Energy architecture of Jiujiu Cabins (2021).}
  The off-grid solar ESS of Jiujiu Cabins is composed of two independent power supply systems combined with the same hybrid ESS.
  This ensures that two PV arrays can generate electricity and supply two sets of 110 V$_\mathrm{AC}$ loads. The ESS is a newly added LFP battery system, which is used in conjunction with the existing lead-acid battery ESS to form a hybrid ESS. An additional backup power input terminal is also added, allowing for the connection of an external generator as a backup power source.
  }
  \label{fig:EnergyArchitecture}
\end{figure}

In order to restore the operation of the energy system, the Hsinchu Forest District Office, Forestry Bureau (renamed as the Forestry and Nature Conservation Agency in 2023) commissioned Super Double Power Technology Co., Ltd. to improve the existing solar energy system of Jiujiu Cabins in 2021~\cite{Project-JiujiuCabinsPVandESS(2020)TW-HsinchuFNCA}. After restructuring and improvement, the off-grid solar energy storage system of Jiujiu Cabins is composed of two independent power supply systems combined with the same ESS architecture, as shown in Fig.~\ref{fig:EnergyArchitecture}. Among them, the first PV array inputs solar energy to hybrid solar inverters, which can transmit power to alternating current (AC) loads. The second PV array can transmit power to another set of AC loads through another hybrid solar inverter. The two sets of hybrid solar inverters share the same hybrid ESS, which consists of a newly established LFP battery~\cite{Sci.Data8(2021)165H.C.Chung} ESS and an existing lead-acid battery ESS. In addition, a new backup power input terminal was added, which can be connected to an external generator as a backup power source.
The details of the energy architecture are described below.

\subsection{Photovoltaic systems (PV systems)}

\begin{figure}[htb]
  \centering
  \includegraphics[width=\columnwidth]{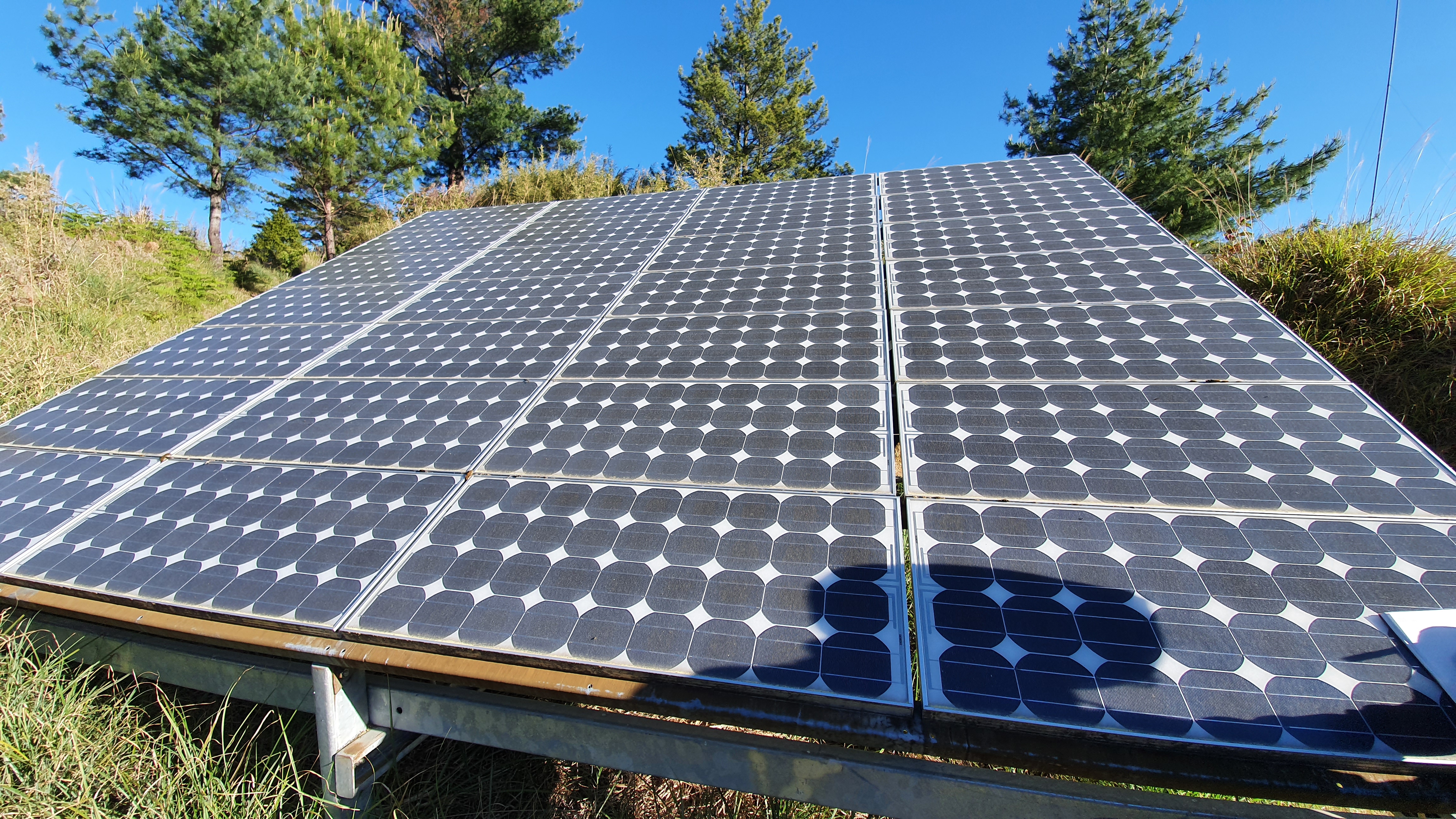}
  \caption{
  \justifying
  \textbf{Ground-mounted PV system 1.}
  PV array 1 is a ground-mounted PV system with a power generation capacity of 2.1 kW$\mathrm{_p}$.
  It is composed of 28 solar modules, each with a capacity of 75 W$\mathrm{_p}$.
  The land area used is approximately 18 m$^2$.
  }
  \label{fig:PV_array_1}
\end{figure}

\begin{figure}[htb]
  \centering
  \includegraphics[width=\columnwidth]{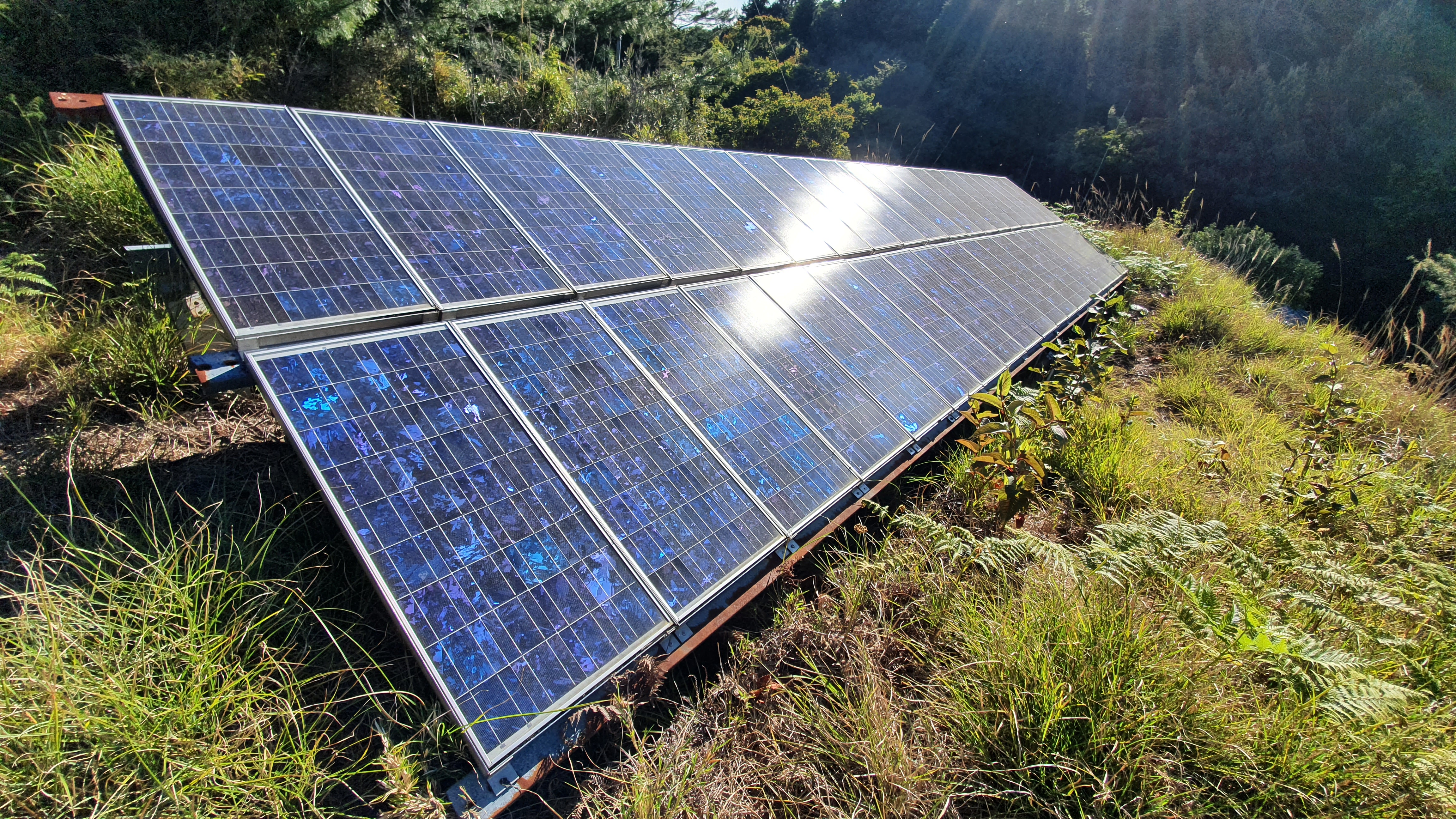}
  \caption{
  \justifying
  \textbf{Ground-mounted PV system 2.}
  PV array 2 is a ground-mounted PV system with a power generation capacity of 3.2 kW$\mathrm{_p}$.
  It is composed of 40 solar modules, each with a capacity of 80 W$\mathrm{_p}$.
  The land area used is approximately 18 m$^2$.
  }
  \label{fig:PV_array_2}
\end{figure}

A photovoltaic (PV) system, as a renewable energy system, converts sunlight directly into direct current (DC) electricity using solar panels.
In Jiujiu Cabins, there are two PV arrays, providing a total power generation capacity of 5.3 kW$\mathrm{_p}$.
The first channel of the PV array is a ground-mounted PV system with a power generation capacity of 2.1 kW$\mathrm{_p}$.
It is composed of 28 solar modules, each with a capacity of 75 W$\mathrm{_p}$ (SIEMENS Solar module SP75), as shown in Fig.~\ref{fig:PV_array_1}.
The second channel of the PV array is a ground-mounted PV system with a power generation capacity of 3.2 kW$\mathrm{_p}$.
It is composed of 40 solar modules, each with a capacity of 80 W$\mathrm{_p}$ (SHARP Solar Module NE-80E2E), as shown in Fig.~\ref{fig:PV_array_2}.
Each of the ground-mounted PV systems occupies approximately 18 m$^2$ (totaling 36 m$^2$).
Although the area is quite small compared to the overall area of the cabins, there are still ecological disturbance issues related to wild animals.


\subsection{Hybrid solar inverters}

The hybrid solar inverter serves the crucial role of linking PV systems, ESS, AC loads, and backup power sources.
The total nominal power of the hybrid solar inverter system is 12 kW.

Some features of the hybrid solar inverters are listed:
(1) detachable liquid crystal display (LCD) control Panel,
(2) built-in Wi-Fi function for APP monitoring,
(3) zero (0 ms) transfer time, the best protection for servers,
(4) pure sine wave inverter with built-in maximum power point tracking (MPPT) solar charger,
(5) selectable power charging current,
(6) configurable AC/Solar input priority via LCD setting,
(7) auto-restart while AC is recovering,
(8) overload and short circuit protection,
(9) cold start function, and
(10) optional parallel operation of up to 9 pcs.

\subsection{LFP energy storage system (ESS)}

ESS serves a role to balance the intermittent renewable energy.
The LFP batteries~\cite{Sci.Data8(2021)165H.C.Chung} are used with BMS in the energy storage system. The batteries are in series connection (16S configuration), making a battery system with a nominal voltage of 48 V$_\mathrm{DC}$ and a capacity of about 10 kWh. A balance circuit is also applied in the system for balancing the voltage of each battery cell. The information of the ESS can be sent to the EMS via the RS-485 communication port.

\subsection{Battery management system (BMS)}

To use the battery safely, a well-designed BMS is required. A BMS is any electronic system that manages a battery cell, pack, or module system, such as protecting the battery from operating outside its safe operating area, monitoring its status, calculating secondary data, reporting that data, controlling its environment, and cell balancing it~\cite{BookChChung2021EngIntePotentialAppOutlooksLiIonBatteryIndustry}.

\subsection{Loads}

Two channels of AC output with a voltage of 110 V$_\mathrm{AC}$ and a frequency of 60 Hz are used in the cabins.
Each channel is provided by one solar hybrid inverter.
Because no communication between the two inverters, their AC output phases are independent.

\subsection{Backup power system}

A channel with a breaker for a backup power system is reserved. Currently, no power system is connected.
A diesel generator could be the backup power system.
When the power from the PV system is insufficient (e.g., rainy days or cloudy days), the diesel generator can supply the insufficient power to the cabins.
When the system encounters failures (e.g., temporary shutdown of the hybrid solar inverter), the diesel generator can supply the full power to the cabins.

\subsection{Energy management system (EMS)}

A basic functionality of an energy management system (EMS) is to monitor and record the energy input and output.
Due to limited funds, the total voltage, current input/output, and power input/output of the ESS are monitored.
The energy information, such as energy generation, consumption, and cell voltages, is not available.
It's difficult to realize the energy input and output of the system quantitatively.
On the other hand, the network communication is not stable.
The signal was intermittent.
However, it can still be a reference for whether a system is functioning.

\begin{figure}[ht]
  \centering
  \includegraphics[width=\columnwidth]{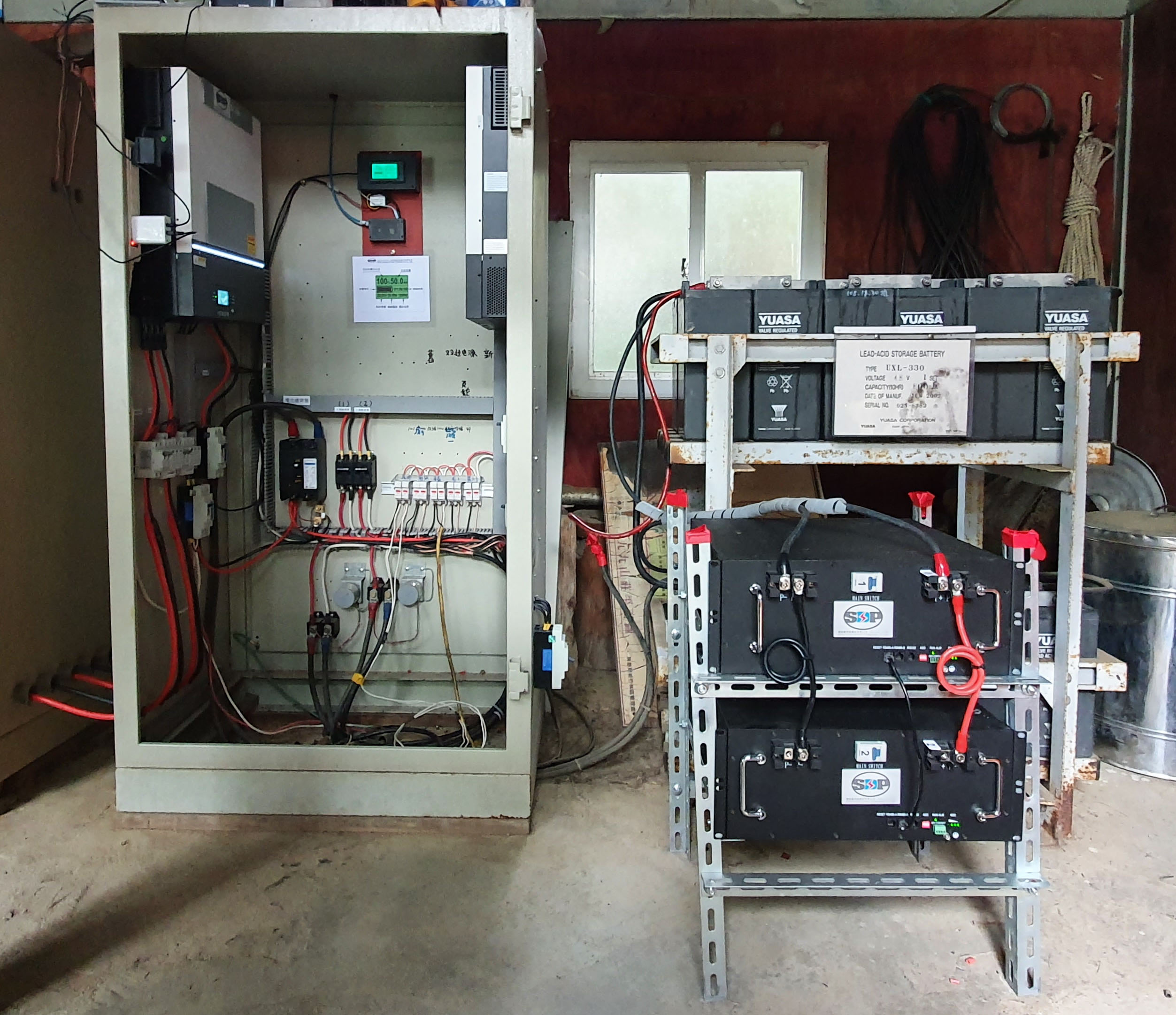}
  \caption{
  \justifying
  \textbf{Machine room of Jiujiu Cabins (2021).}
  }
  \label{fig:MachineRoom}
\end{figure}

\section{Historical development of energy systems}
\label{sec:HistoricalDevelopmentEnergySystems}

According to data from the Government e-Procurement System~\cite{GovernmentE-ProcurementSystem()PCC}, although the Hsinchu Forest District Office commissioned Sheng Yuan Construction Co., Ltd. to carry out a solar panel improvement project in 2003~\cite{Project-JiujiuCabinsToiletAndPV(2003)TW-HsinchuFNCA}, the system may have undergone several changes over the past decade, and it is impossible to know which companies handled it. This has led to the system becoming complex and difficult to maintain.

The Jiujiu Cabins had two existing electrical systems, designated as Old System 1 and Old System 2. These systems were likely installed by different manufacturers at different times. Owing to a malfunction in the Old System 1, power was unavailable. Another manufacturer then installed Old System 2 to supply power to the cabins' 110 V$_\mathrm{AC}$ loads. However, the inverter in Old System 2 also failed due to aging, causing power supply anomalies and no outputting electricity.

In order to restore the operation of the energy system, the Hsinchu Forest District Office commissioned Super Double Power Technology Co., Ltd. to reorganize, repair, and improve the existing solar energy system of Jiujiu Cabins in 2021~\cite{Project-JiujiuCabinsPVandESS(2020)TW-HsinchuFNCA}.
After the improvement project, it's the current energy architecture (which is described in Sec.~\ref{sec:ArchitectureEnergySystem} in detail).
The evolution of the energy system in Jiujiu Cabins is shown in Table~\ref{tab:EnergySystemEvolution}.

\begin{table*}[htb]
  \centering
  \begin{tabular}{| c | c | c | c | c | c |}
    \hline
    \textbf{Year} & \textbf{Renewable energy} & \textbf{Energy storage system} & \textbf{Load capacity} & \textbf{Energy management system} & \textbf{Project} \\
    \hline
    2021 & Total: 5.3 kW$\mathrm{_p}$ (PV) & Total: 25 kWh (Hybrid) & 12 kW & Yes & \cite{Project-JiujiuCabinsPVandESS(2020)TW-HsinchuFNCA} \\
     &  &  &  &  &  \\
     & Component: & Component: &  &  &  \\
     & Ch 1: 2.1 kW$\mathrm{_p}$ (PV) & New LFP: 10 kWh &  &  &  \\
     & Ch 2: 3.2 kW$\mathrm{_p}$ (PV) & Old lead-acid: 15 kWh &  &  &  \\
    \hline
    Before & Total: 5.3 kW$\mathrm{_p}$ (PV) & 15 kWh (Lead-acid) & 3 kW & No &  \\
    2021 &  &  &  &  &  \\
     & Component: &  &  &  &  \\
     & Ch 1: 2.1 kW$\mathrm{_p}$ (PV) &  &  &  &  \\
     & Ch 2: 3.2 kW$\mathrm{_p}$ (PV) &  &  &  &  \\
    \hline
  \end{tabular}
  \caption{Evolution of the energy system of Jiujiu Cabins}
  \label{tab:EnergySystemEvolution}
\end{table*}

\section{On-site survey of energy systems}
\label{sec:On-siteSurveyEnergySystems}

The aging of the energy system and lack of maintenance from original system integrators caused multiple system failures. The Forestry Bureau technician explained that although there were system integrators involved in planning and construction, after the warranty expired, the original design and construction companies stopped providing support, leaving the work to be handled by local small plumbing companies.

Most small-scale plumbers are experts in AC power systems or general household plumbing, with limited knowledge of solar DC systems and experience in integrating AC/DC systems. They can often only replace components or equipment based on the original system, lacking the background knowledge to adjust parameter settings to ensure stable system operation. Furthermore, the selected equipment sometimes differs from the original system provider's, causing problems.

In order to understand the energy system status of Jiujiu Cabins, from May 12 to 13, 2021, the Hsinchu Forest District Office, ``Wei-Chih, CHEN Architect \& Associates,'' and Super Double Power Technology Co., Ltd. conducted an on-site survey at Jiujiu Cabins.

\begin{figure}[htb]
  \centering
  \includegraphics[width=\columnwidth]{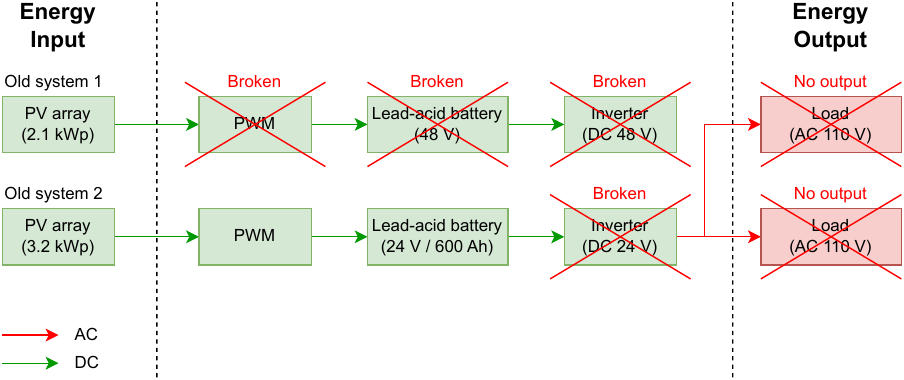}
  \caption{
  \justifying
  \textbf{Old energy architecture of Jiujiu Cabins (2021-05-13).}
  The solar charger, lead-acid battery, and inverter of the Old System 1 all failed. Therefore, the Old System 2 was used instead, but the inverter of the Old System 2 also failed, resulting in a complete power supply abnormality and inability to output power.
  }
  \label{fig:EnergyArchitecture-Old}
\end{figure}

The Jiujiu Cabins has two power systems, namely Old System 1 and Old System 2, as shown in Fig.~\ref{fig:EnergyArchitecture-Old}. These two systems were likely installed by different manufacturers at different times.

The Old System 1 can be considered as a microgrid (MG) system built on a 48 V$_\mathrm{DC}$ DC bus. Solar panels (2.1 kW$\mathrm{_p}$) generate electricity, which is then charged via a solar charger (PWM) to a 48 V$_\mathrm{DC}$ lead-acid battery ESS. A 48 V$_\mathrm{DC}$ inverter converts the DC power to 110 V$_\mathrm{AC}$ for output. Based on the property label, its installation date is Jan. 20, 2000.

The Old System 2 can be considered as a MG system built on a 24 V$_\mathrm{DC}$ DC bus. After generating electricity from the solar panels (3.2 kW$\mathrm{_p}$), the electricity is charged to the 24 V$_\mathrm{DC}$ lead-acid battery ESS through a solar charger (PWM). The DC power is then converted to 110 V$_\mathrm{AC}$ AC power output by a 24 V$_\mathrm{DC}$ inverter, with a rated power of 3 kW.

\noindent \textbf{Fault Status of Old System 1:}
\begin{enumerate}
  \item 48 V$_\mathrm{DC}$ lead-acid battery ESS: The batteries are old and faulty. Each 12 V$_\mathrm{DC}$ lead-acid battery cell has an open-circuit voltage of 0.7 V$_\mathrm{DC}$$\sim$5.6 V$_\mathrm{DC}$, both far below the normal discharge cutoff voltage of 10 V$_\mathrm{DC}$, indicating battery failure.
  \item 48 V$_\mathrm{DC}$ inverter failure: Unable to output 110 V$_\mathrm{AC}$ power.
  \item Solar charger failure.
\end{enumerate}

\noindent \textbf{Fault Status of Old System 2:}
\begin{enumerate}
  \item The 24 V$_\mathrm{DC}$ inverter is faulty and unable to output 110 V$_\mathrm{AC}$ power.
  \item A bypass line connecting the PWM and the inverter (not shown in the figure). While this line could provide power, it would cause the inverter to start and stop intermittently when solar energy was weak. After a period of operation, this would damage the internal electronic components and cause equipment failure. This is a poor system circuit design.
\end{enumerate}

\section{Solar power system improvement project (2021)}
\label{sec:SolarPowerSystemImprovementProject2021}

In 2021, the Hsinchu Forest District Office commissioned Super Double Power Technology Co., Ltd. to carry out a solar energy system improvement project~\cite{Project-JiujiuCabinsPVandESS(2020)TW-HsinchuFNCA} to solve the system failure issues.
The project took place in August 2021.

\subsection{Microgrid types and energy storage system configuration}

Microgrid (MG) topologies can be classified into three main categories: AC microgrids (AC-MG), DC microgrids (DC-MG), and hybrid microgrids (Hybrid-MG)~\cite{Renew.Sustain.EnergyRev.24(2013)387J.J.Justo, Renew.Sustain.EnergyRev.17(2013)147E.Planas, EnergyConvers.Manag.86(2014)132E.Hossain}.

AC-MG are the most commonly used configuration because they provide a way to directly integrate distributed generation units (DG units) into the existing public grid with only minor modifications. This architecture is characterized by easy voltage level regulation via low-frequency transformers and strong fault management capabilities (e.g., fault detection and handling through multiple protection devices). However, it also has some drawbacks, such as the need to synchronize distributed generation units or reactive power circulation, which increases power losses in the transmission system. The feasibility of AC-MG integration has been extensively studied in numerous publications~\cite{Renew.Sustain.EnergyRev.15(2011)186N.W.A.Lidula, EnergyConvers.Manag.86(2014)132E.Hossain, Renew.Sustain.EnergyRev.40(2014)659M.Soshinskaya}.

Even though most distribution networks are powered by AC, the high penetration rate of DC DG units, ESSs, and loads, along with other characteristics, promotes the development of DC-powered distribution networks. Its main advantages lie in higher overall energy efficiency due to the use of fewer interface converters, reducing conversion losses, and the absence of reactive current circulation in the network. Furthermore, synchronous DG is not required. However, this configuration requires large-scale upgrades to existing distribution networks, thus significantly increasing costs. Several studies have demonstrated that DC distribution configurations still have advantages~\cite{Renew.Sustain.EnergyRev.24(2013)387J.J.Justo, Renew.Sustain.EnergyRev.17(2013)147E.Planas, Renew.Sustain.EnergyRev.43(2015)415I.Patrao, Renew.Sustain.EnergyRev.43(2015)726E.Planas}.

Hybrid AC/DC MG configurations have attracted significant attention due to their integration of the advantages of both AC and DC architectures~\cite{Renew.Sustain.EnergyRev.43(2015)415I.Patrao, Renew.Sustain.EnergyRev.43(2015)726E.Planas, IEEEJ.Emerg.Sel.Top.PowerElectron.6(2018)738S.K.Sahoo, Appl.Sci.11(2021)6242O.Azeem}. Their key feature is the integration of AC and DC networks into a single distribution network, facilitating direct access to distributed AC and DC power sources, energy storage systems, and loads. This characteristic provides an efficient way to connect upcoming renewable energy (RES) or electric vehicle (EV) units with minimal modifications to the existing distribution network, thus reducing overall costs. While hybrid AC/DC MGs represent an excellent solution for integrating smart grids into traditional distribution networks, they remain under development because most research focuses on AC or DC systems separately.

MGs with ESSs have become a promising component in smart grid deployments~\cite{IEEETrans.Industr.Inform.10(2014)152M.L.D.Silvestre, IEEETrans.Industr.Inform.62(2015)2540G.Graditi}. However, due to the intermittent nature of renewable energy and the instability of load demand, the electricity in MGs sometimes cannot alleviate instantaneous or short-term load demands, leading to system frequency fluctuations~\cite{IntJ.Electr.PowerEnergySyst.54(2014)325M.R.Aghamohammadi}. In such cases, ESSs must be utilized to smooth out fluctuating renewable energy to provide high power quality~\cite{IEEETrans.Ind.Electron.60(2013)1254J.M.Guerrero, IEEETrans.Ind.Electron.60(2013)1263J.M.Guerrero, IEEETrans.EnergyConvers.29(2014)204F.Diaz-Gonzalez}.

Aggregated ESS and distributed ESS are two basic configurations of ESSs used in MG applications. Aggregated ESS serves the entire MG, while distributed ESS is deployed near the power generation unit~\cite{ComparisonESSTechnologiesConfigurationsWindFarm(2007)1280}. The capacity of the ESS can be used to mitigate power fluctuations~\cite{ComparisonESSTechnologiesConfigurationsWindFarm(2007)1280}. As the capacity of the ESS increases, the cost also increases. Large-scale ESSs are difficult to manufacture and control. Therefore, small distributed ESSs can be used to achieve reliable and efficient power regulation. In distributed energy storage configurations, the energy storage units are directly connected to specific distributed power sources through numerous interfaces. However, power control is a major challenge for distributed systems. Furthermore, there is still some energy loss during the storage process, including conversion losses through the power electronic interfaces between the distributed power source and the ESS, as well as electrochemical cell conversion losses~\cite{IEEEAccess6(2018)35143M.Faisal}.

In Jiujiu Cabins, the Old System 1 and 2 are treated as AC-MG architectures (since the systems outputting power to loads via a 110 V$_\mathrm{AC}$ AC bus). In this improvement project, the entire system continued to utilize the AC-MG architecture, while the ESS configuration transitioned from a distributed configuration to an aggregated one.

\subsection{Key issues}

The following are the key issues that need to be addressed:
\begin{enumerate}
  \item Interference may occur between the two old systems under certain operating conditions, which is one of the reasons why the systems are prone to failure later. The two old systems need to be reassembled into one system.
  \item The original inverters are all damaged and need to be replaced.
  \item The two existing PV arrays can be reused, but the series and parallel connections need to be adjusted within the inverter specifications.
  \item The original lead-acid battery ESS is aging and has insufficient capacity, requiring an increase in the ESS's capacity to ensure nighttime power supply.
  \item The power lines need to be rebuilt.
\end{enumerate}

\subsection{Main engineering subprojects}

The solution involves restructuring the solar power lines, upgrading the inverters, restructuring the power supply lines at the cabins, and increasing the capacity of the ESS.
However, this is not a simple matter of adding more batteries in parallel. It involves the parallel connection of old and new ESSs and the integration of the entire energy system.

The main engineering subprojects are: (1) adding LFP battery modules, (2) updating inverters, (3) modifying and reorganizing the existing PV array 1, (4) modifying and reorganizing the existing PV array 2, (5) changing control boxes (changing AC/DC lines), and (6) materials (hardware parts/connectors/DC lines). The following describes each subproject in detail:
\begin{enumerate}
  \item \textbf{Adding LFP battery modules.} The original lead-acid battery pack was 24 V$_\mathrm{DC}$ / 600 Ah. After being modified by connecting it in series and parallel, it was reassembled into a 48 V$_\mathrm{DC}$ / 300 Ah battery pack with a total capacity of approximately 15 kWh. Two new 48 V$_\mathrm{DC}$ / 100 Ah LFP batteries were added, with a total capacity of approximately 10 kWh. The modified lead-acid battery pack (15 kWh) and the new LFP battery pack (10 kWh) were connected in parallel, for a total capacity of approximately 25 kWh (neglecting the aging of the lead-acid battery). If we use a distributed ESS configuration, a specific power conversion system (PCS) or DC/DC converter should be used as a physical separation device connecting to the main bus. In this project, we used an aggregated ESS configuration in parallel connection of the LFP and lead-acid ESSs. As their nominal voltages are close, the physical separation device between the two ESSs could be omitted, while the operation voltage range should be adjusted to the intersection of the two ESSs' voltage ranges. This work could be achieved by adjusting the operation voltage range in the BMS.
  \item \textbf{Updating inverters.} The damaged inverter (24 V$_\mathrm{DC}$ / 3000 W) was removed. Two new independent hybrid solar inverters (48 V$_\mathrm{DC}$ / 6000 W) were installed. After installation, the total output AC power is 12 kW.
  \item \textbf{Modifying and reorganizing the existing PV array 1.} The PV series and parallel configuration of PV array 1 (2.1 kW$_\mathrm{p}$) was modified, so that its operating voltage is in the range of 90$\sim$230 V$_\mathrm{DC}$ and its open circuit voltage (OCV) is less than 250 V$_\mathrm{DC}$, which can meet the inverter input specifications.
  \item \textbf{Modifying and reorganizing the existing PV array 2.} The PV series and parallel configuration of PV array 2 (3.2 kW$_\mathrm{p}$) was modified, so that its operating voltage is in the range of 90$\sim$230 V$_\mathrm{DC}$ and its OCV is less than 250 V$_\mathrm{DC}$, which can meet the inverter input specifications.
  \item \textbf{Changing control boxes (changing AC/DC lines).}
  (a) Two damaged solar chargers were removed;
  (b) the damaged inverter was removed;
  (c) the AC/DC circuit was modified; and
  (d) a generator circuit was added.
  \item \textbf{Materials (hardware parts, connectors, and DC lines).}
  (a) The old hardware parts, connectors, and DC lines were removed;
  (b) new hardware parts, connectors, DC lines were added according to the current decoration situation, including DC surge absorbers, DC dedicated leakage circuit breakers, coulomb meters, cloud EMS, and generator charging input terminals.
\end{enumerate}

\subsection{Estimation of energy consumption, storage, and generation}
\label{sec:EstimationEnergyConsumption}

Regarding energy consumption, daily electricity consumption $E^{day}$ can be assessed using the following formula.
\begin{equation}
E^{day} = \sum_{i=1}^{n} \int_{0}^{1~day} P_i (t) dt,
\label{eq:EnergyConsumption}
\end{equation}
where $P_i (t)$ is the function of the power consumption of the $i$-th device as a function of time.

After observing the electrical equipment and investigating usage habits at Jiujiu Cabins, an analysis of the equipment and electricity consumption was conducted, as shown in Table~\ref{tab:DailyElectricityConsumption}. Since the most frequently used equipment in the cabins is lighting, its usage time is approximately 17:00 $\sim$ 20:00 and 03:00 $\sim$ 06:00 (about 6 hours per day).

\begin{table*}[htb]
  \centering
  \begin{tabular}{| c | l | l | r | r | r | r | r |}
    \hline
    \textbf{No.} & \textbf{Place} & \textbf{Equipment} & \textbf{Power} & \textbf{Time} & \textbf{Number of} & \textbf{Daily usage} & \textbf{Daily power} \\
     &  &  & \textbf{(kW)} & \textbf{use (h)} & \textbf{uses per day} & \textbf{time (h)} & \textbf{consumption (kWh)} \\
    \hline
        & & & \textbf{A} & \textbf{B} & \textbf{C} & \textbf{D = B $\times$ C} & \textbf{E = A $\times$ D} \\
    \hline
    1 & Restaurant       & 20W energy-saving light bulb $\times$ 2 & 0.040 &  3 & 2 &  6 & 0.24 \\
      & (reception room) & 20W energy-saving light bulb $\times$ 2 & 0.040 &  3 & 2 &  6 & 0.24 \\
    2 & Chengkung Yurts             & 20W energy-saving light bulb $\times$ 6 & 0.120 &  3 & 2 &  6 & 0.72 \\
      & & (Each of the Chengkung Yurts has 1 lamp.)  &  &   &  &   & \\
    3 & Longmen Inns                & 20W energy-saving light bulb $\times$ 6 & 0.120 &  3 & 2 &  6 & 0.72 \\
      & & (Each of the Longmen Inns has 2 lamps.)    &  &   &  &   & \\
    4 & Toilet                      & 20W energy-saving light bulb $\times$ 1 & 0.020 &  3 & 2 &  6 & 0.12 \\
    5 & Machine room                & 20W energy-saving light bulb $\times$ 1 & 0.020 &  3 & 2 &  6 & 0.12 \\
      &                             & Hybrid inverter                         & 0.070 & 24 & 1 & 24 & 1.68 \\
    \hline
      &                             &                                         &       &    &   &    & \textbf{Sum: 3.60} \\
    \hline
  \end{tabular}
  \caption{Daily electricity consumption estimation for Jiujiu Cabins}
  \label{tab:DailyElectricityConsumption}
\end{table*}

The energy storage capacity is related to how much energy resilience the system can provide to the mountain hut, that is, how long the system can last when there is no solar power generation. During peak hiking season, when there are many hikers (the cabins can accommodate 150 people), and both Chengkung Yurts and Longmen Inns are fully booked, the daily electricity consumption can reach 3.6 kWh. Without any solar power, the system can last approximately 6 days (25 kWh / 3.6 kWh/day = 6.9 days). During the off-season, when there are fewer hikers (fewer than 50 people), only Chengkung Yurts are used, and the daily electricity consumption can reach 2.88 kWh. Without any solar power, the system can last approximately 8 days (25 kWh / 2.88 kWh/day = 8.6 days). During the mountain closure period, and only the administrator is present, the daily electricity consumption can reach 1.92 kWh. Without any solar power, the system can last approximately 13 days (25 kWh / 1.92 kWh/day = 13.0 days). Diesel generators are only needed to supplement power when the ESS is at extremely low energy levels and solar power generation is low. However, considering the cloudy rainy season, the mountain will be closed, and climbing activities will be prohibited. At this time, there will only be a need to use the generator if there is insufficient sunlight for more than ten consecutive days, so the generator is not expected to be used very often.

Regarding solar power generation, the current solar installation capacity is 5.3 kWp. Assuming maximum power generation for 3 hours per day, the daily energy generation could reach approximately 15.9 kWh (5.3 kW × 3 h = 15.9 kWh). This solar power generation is significantly higher than the cabins' maximum daily energy consumption of 3.6 kWh, ensuring the ESS is always fully charged. However, on cloudy days with insufficient sunlight, when maximum power generation time is less than 40 minutes, the daily energy generation could reach approximately 3.5 kWh (5.3 kW × 40/60 h = 3.5 kWh), falling below the maximum daily energy consumption. This causes the ESS to be unable to maintain a full charge, and the remaining capacity will gradually decrease. If the ESS is at extremely low remaining capacity, with good weather and sufficient sunlight, it will take approximately 1.5 days of full-power PV generation to restore the ESS to a full charge state (25 kWh / 15.9 kWh/day = 1.57 days).

\subsection{EMS data analysis}
\label{sec:EMSDataAnalysis}

\begin{figure}[htb]
  \centering
  \includegraphics[width=\columnwidth]{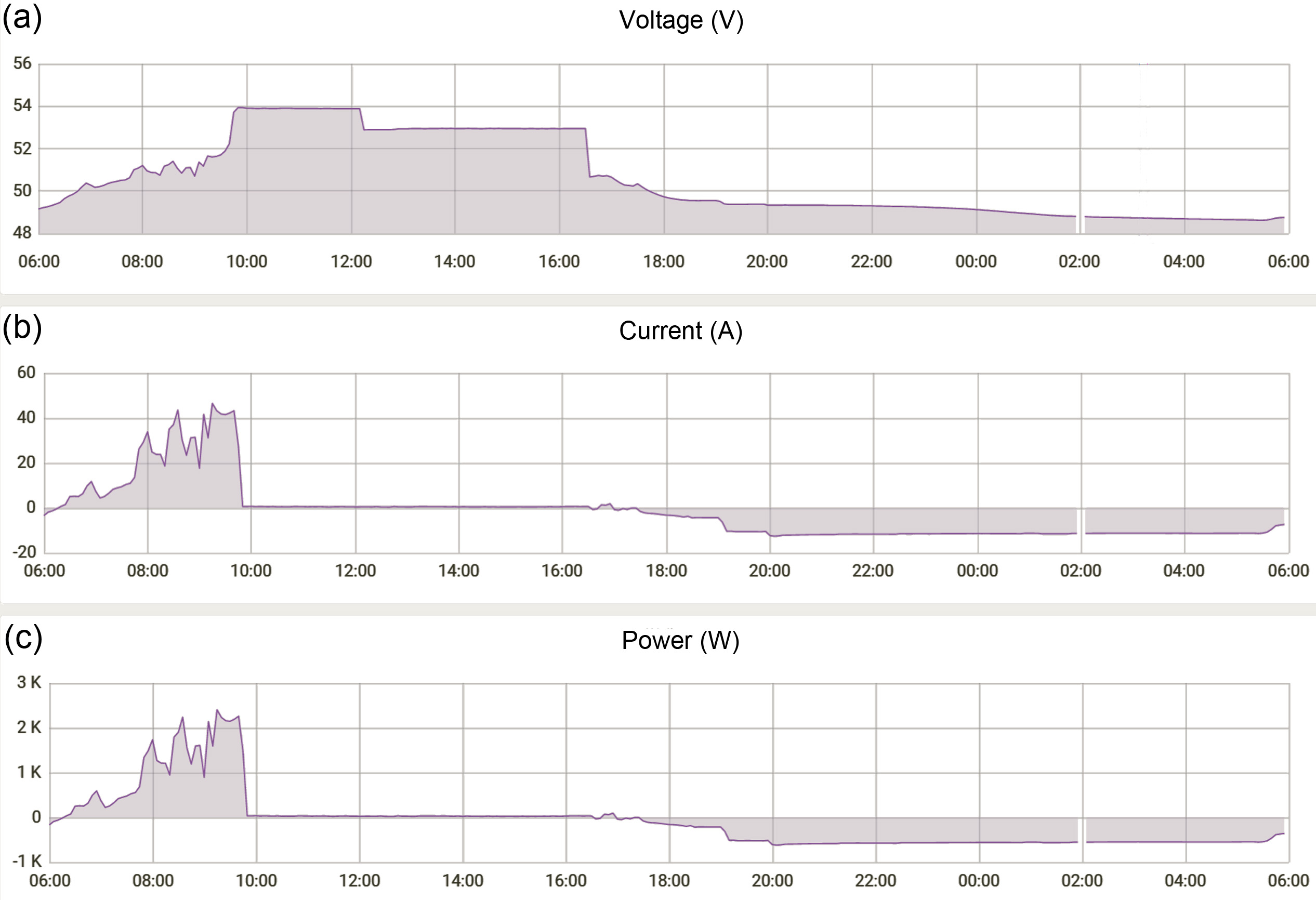}
  \caption{
  \justifying
  \textbf{24-hour charge and discharge data of ESS.}
  (a) Voltage variation, (b) current variation, and (c) power variation.
  Date and time: 2021-08-18 06:00 $\sim$ 2021-08-19 06:00.
  }
  \label{fig:EMS_data_2021}
\end{figure}

In this project, due to budget constraints, the self-developed cloud-native EMS only records the total voltage, input/output current, and input/output power of the ESS. Because the cabins' energy usage has a certain cyclical characteristic on a daily basis, a one-day data analysis is used to assess the system's condition, as shown in Fig.~\ref{fig:EMS_data_2021}.

The 24-hour data shows that the highest total voltage of the ESS does not exceed 54 V$_\mathrm{DC}$ (as shown in Fig.~\ref{fig:EMS_data_2021}(a)), indicating that the BMS provides stable control over the LFP battery, keeping it within a safe operating voltage range (the rated voltage of this ESS is 48 V$_\mathrm{DC}$).

Around 6:00 AM, a while after sunrise, the PV systems charge the ESS, and the current display shows a positive value as shown in Fig.~\ref{fig:EMS_data_2021}(b), indicating charging (a negative current indicates discharging). As the sunlight gradually intensifies, the charging current increases, reaching a maximum of 46 A$_\mathrm{DC}$ around 9:15 AM. However, after 9:40 AM, because the ESS is fully charged, the system stops charging, and the current can be seen to drop instantly to 0 A$_\mathrm{DC}$. At this point, the cabins' electricity is entirely supplied by solar energy.

Around 5 PM, solar power generation weakened to the point where it was insufficient to supply the cabins' electricity. The shortfall was covered by the ESS, and the discharge current of the ESS could be seen gradually increasing. By around 7 PM, solar power had completely stopped supplying electricity, and the ESS was fully engaged to power the cabins, resulting in a noticeable drop in the discharge current. The cabins' electricity consumption peaked around 8 PM, at approximately 620 W (as shown in Fig.~\ref{fig:EMS_data_2021}(c)), and continued until nearly 6 AM the next morning when solar power began generating electricity to support part of the cabins' needs, at which point the discharge current of the energy storage system gradually decreased. The above is a 24-hour electricity consumption cycle record and analysis of the Jiujiu Cabins, demonstrating the stable operation of the system.

\subsection{System benefit verification}

Based on the periodic changes in the ESS data (see Sec.~\ref{sec:EMSDataAnalysis}), the system is operating stably. Solar power generation far exceeds daily power consumption and replenishes the ESS’s nighttime power consumption, keeping it at full charge and providing the cabins with sufficient energy resilience (see Sec.~\ref{sec:EstimationEnergyConsumption}). Regarding equipment reliability, the system underwent only routine maintenance during the three-year warranty period, with no serious equipment damage (e.g., replacing the inverter and causing a system shutdown for a long time). Although the equipment is out of warranty, it is still operational. This confirms that the equipment operates well in the mountains and can meet energy needs through renewable energy. This system transforms the role of traditional diesel generators from a primary power source to a supplementary one, significantly reducing diesel consumption and meeting the environmental protection requirements of high-altitude areas.

\section{Future plan for the energy system}
\label{sec:FuturePlanningForEnergySystem}

After completing the aforementioned improvement project in 2021, architect Wei-Chih Chen asked Super Double Power Technology to design the future energy system of Jiujiu Cabins. The system is expected to provide greater power to loads such as medical equipment, communication equipment, and lighting equipment. On the other hand, greater energy storage capacity will address the intermittent nature of renewable energy sources, improving system resilience.

This plan can be divided into four main components: (1) replacing the old PV arrays with a 20 kW$_\mathrm{p}$ PV system, (2) replacing the environmentally unfriendly lead-acid battery ESS (15 kWh) with the original LFP ESS (10 kWh), and adding an additional 80 kWh of LFP ESS, (3) installing a self-developed cloud-native EMS, and (4) integrating AC and DC systems. They are stated in detail below.

\subsection{PV system replacement}

Key engineering points are listed:
\begin{enumerate}
  \item The existing PV modules of low efficiency (two channels, totaling 5.3 kW$_\mathrm{p}$) should be dismantled and removed. Existing usable brackets and fasteners should be collected and retained.
  \item New PV arrays should have a capacity of 20 kW$_\mathrm{p}$ or higher, and the PV modules should be Class-A with a module efficiency of 20\% or higher. Product verification conforming to IEC performance and safety standards should be obtained.
  \item Solar DC power protection devices such as surge absorbers, lightning bypass, short-circuit protectors, and overload circuit breakers should be installed in the solar DC power control enclosure.
  \item Existing aluminum PV brackets should be assessed for any corrosion, paint peeling, cracking, deformation, or other conditions affecting structural strength, and a design should be implemented to maximize their retention and reuse to achieve environmental protection goals.
  \item Outdoor connections to PV power lines should have sufficient high-voltage insulation and waterproofing capabilities. For example, the St\"{a}ubli MC4 solar-specific waterproof connector can be used, which has 1000 V high voltage and IP68 waterproofing capabilities.
  \item To withstand the extremely harsh climate of high-altitude, low-temperature environments, the silicone material used for filling gaps should possess excellent weather resistance, maintaining its elasticity for extended periods. It should be ozone-resistant, ultra-violet-resistant (UV-resistant), and free from aging, cracking, or damage. It should be moisture-resistant and corrosion-resistant, preventing severe electrolytic corrosion. It should withstand temperature variations between freezing and boiling points while retaining the properties of the raw materials. It should have good adhesion and remain bonded for a long time without detaching.
  \item During construction, a diesel generator paired with an ESS can be used to supply electricity to the cabins, ensuring a normal power supply of the energy system.
\end{enumerate}

\subsection{ESS replacement and enlargement}

To accommodate the increased demand for electrical equipment (such as AED (Automated External Defibrillator), 4G/5G base station equipment, and lighting equipment), a larger ESS is needed to sustain operations during periods of insufficient renewable energy generation. 4G/5G base station equipment, operating 24/7, consumes an average of 12 kWh per day. Combined with the existing regular power consumption of the cabins, this brings the average daily consumption to approximately 15 kWh. Therefore, the overall ESS needs to be upgraded to over 90 kWh to maintain the original system's design for approximately 6 days. In addition, given the limited space in the machine room, a device with 90 kWh of electricity occupies about two standard server racks, which is a reasonable choice.

Key engineering points are listed:
\begin{enumerate}
  \item The existing ESS comprises two groups: one is a 15 kWh lead-acid battery ESS, which is to be phased out due to battery aging and poor energy storage performance; the other is a 10 kWh LFP ESS built in 2021, which will be retained and integrated with the newly purchased energy storage equipment in this plan.
  \item The newly constructed ESS uses an 80 kWh LFP ESS with a rated voltage of 48 V$_\mathrm{DC}$ and integrates with the operation of the existing ESS.
  \item The ESS cabinet can adopt a modular design to distribute the cabinet weight. This allows the mountain porters to successfully bring the materials up the mountain.
  \item The battery can adopt a modular design to distribute the battery weight. This allows the mountain porters to successfully bring the materials up the mountain.
\end{enumerate}

\subsection{EMS development}

The new energy system is expected to have approximately 40 energy nodes. An embedded system will collect, calculate, statistically analyze, and store data from these nodes. All energy data will be uploaded to a cloud-based energy management system via an industrial-grade 4G LTE router.

This new EMS offers energy management through the integration of hardware and software. Its core functions can be summarized in two main aspects: (1) Energy Consumption Monitoring: Real-time collection of all energy input and output data, displayed on dashboards showing the power consumption status of each area or device. (2) Data Analysis: Generating energy consumption reports, analyzing historical trends, and identifying ``energy-saving hotspots'' or abnormal energy consumption behaviors. For example, in a mountain hut, it can obtain real-time solar power generation and daily accumulative power generation, as well as real-time and daily accumulative electricity consumption for each location. After a full year of data, it enables an understanding of the annual energy output and usage, assesses the adequacy of solar panel capacity and ESS capacity, and subsequently initiates energy planning. For instance, during peak energy usage periods, sufficient diesel fuel needs to be stockpiled to maintain the energy system's operational capacity.

This plan will integrate hardware and software while preserving the legacy system architecture to the greatest extent possible. The main tasks of the cloud-based EMS are:
\begin{enumerate}
  \item Hardware integration of the ground system
  \item Software integration of the ground system
  \item Software integration of the cloud system
\end{enumerate}
The three main work projects are explained as follows.

\subsubsection{Hardware integration of the ground system}

The points to note are listed below:
\begin{itemize}
  \item Install 7 smart DC meters (solar + DC load). Meter accuracy should be Class-0.5 or higher.
  \item Install 3 smart AC meters (220 V$_\mathrm{AC}$ output). Meter accuracy should be Class-0.5 or higher.
  \item Install 1 smart AC meter (110 V$_\mathrm{AC}$ output). Meter accuracy should be Class-0.5 or higher.
  \item Add 4 industrial-grade embedded systems (for input/output). Hardware-wise, it should have an industrial-grade storage temperature range (-40$\sim$80$^\circ$C) and an industrial-grade operating temperature range (-20$\sim$60$^\circ$C) to cope with high-altitude climate changes, and possess communication interfaces such as Ethernet, UART, RS-485, RS-232, and CAN Bus. Software-wise, it adopts a Linux-based operating system architecture, capable of decoding and performing input/output operations for Ethernet, UART, RS-485, RS-232, and CAN Bus communications. For storage, it uses industrial-grade, high-durability SD cards.
  \item Install 4 industrial-grade embedded systems (for ESS). Requirements are as stated in the previous point.
  \item Add a new router.
  \item Reorganize network, RS-485, RS-232, and CAN Bus communication lines.
  \item Reorganize the power lines for communication devices.
\end{itemize}

\subsubsection{Software integration of the ground system}

The points to note are listed below:
\begin{itemize}
  \item For the 40 newly added energy nodes, integrate ground-level energy information, including voltage, current, power, etc. (information listed below).
  \item Decode the new protocol of ESS and calculate energy data, and develop embedded system software for computation.
\end{itemize}
The energy information from the input and output terminals is listed below:
\begin{itemize}
  \item PV input data: voltage, current, power, and total energy generation.
  \item 220 V$_\mathrm{AC}$ input/output data: voltage, current, power, power factor (PF), total energy generation/consumption.
  \item 110 V$_\mathrm{AC}$ output data: voltage, current, power, PF, total energy consumption.
\end{itemize}
Energy statistics items are listed below:
\begin{itemize}
  \item Daily total PV energy generation, daily total energy consumption of the cabins, and daily energy generation of the generator.
  \item Monthly total PV energy generation, monthly total energy consumption of the cabins, and monthly energy generation of the generator.
  \item Annual total PV energy generation, annual total energy consumption of the cabins, and annual energy generation of the generator.
\end{itemize}

\subsubsection{Software integration of the cloud system}

To understand the energy status of the system, it is essential to be able to grasp the energy input and output. For input, please refer to the energy input line monitoring settings as shown in Table~\ref{tab:EMS_Input}; for output, please refer to the energy output line monitoring settings as shown in Table~\ref{tab:EMS_Output}.
Several notes related to the software integration of the cloud system are listed in the appendix section.

\begin{table}[htb]
  \centering
  \begin{tabular}{| c | l | l | r |}
    \hline
    \textbf{No.} & \textbf{Line name} & \textbf{Description} & \textbf{Nominal voltage} \\
    \hline
    1 & PV1 & PV array 1 & $<600$ V$_\mathrm{DC}$ \\
    \hline
    2 & PV2 & PV array 2 & $<600$ V$_\mathrm{DC}$ \\
    \hline
    3 & PV3 & PV array 3 & $<600$ V$_\mathrm{DC}$ \\
    \hline
    4 & PV4 & PV array 4 & $<600$ V$_\mathrm{DC}$ \\
    \hline
    5 & PV5 & PV array 5 & $<600$ V$_\mathrm{DC}$ \\
    \hline
    6 & PV6 & PV array 6 & $<600$ V$_\mathrm{DC}$ \\
    \hline
    7 & Generator & Diesel generator & 220 V$_\mathrm{AC}$ \\
    \hline
  \end{tabular}
  \caption{Energy input line monitoring settings}
  \label{tab:EMS_Input}
\end{table}

\begin{table}[htb]
  \centering
  \begin{tabular}{| c | l | p{27mm} | r |}
    \hline
    \textbf{No.} & \textbf{Line name} & \textbf{Description} & \textbf{Nominal voltage} \\
    \hline
    1 & LineGeneral & Typical 110 V$_\mathrm{AC}$ power consumption & 110 V$_\mathrm{AC}$ \\
    \hline
    2 & Line220VAC & Dedicated for 220 V$_\mathrm{AC}$ loads & 220 V$_\mathrm{AC}$ \\
    \hline
    3 & Line48VDC & Dedicated for 48 V$_\mathrm{DC}$ loads & 48 V$_\mathrm{DC}$ \\
    \hline
    4 & LineEMS & Monitoring the power consumption of the EMS & 220 V$_\mathrm{AC}$ \\
    \hline
  \end{tabular}
  \caption{Energy output line monitoring settings}
  \label{tab:EMS_Output}
\end{table}

\subsection{AC and DC systems integration}

This project involves both AC and DC systems within a MG. The power lines for these two systems should be independent; otherwise, system damage may occur. Appropriate physical isolation (e.g., breakers, switches, etc.) is required between each device to facilitate installation and subsequent maintenance. Other points to note:
\begin{enumerate}
  \item The PV system, ESS, and EMS mentioned above need to be integrated, with parameters matched to ensure they can operate collaboratively.
  \item The existing inverters should be used in conjunction with the newly installed inverters, operating in parallel to supply power to the cabins.
  \item The inverter's DC operating voltage setting should match the operating voltage setting of the LFP ESS to prevent the inverter from draining the battery's power.
  \item When connecting different batteries in parallel, the system voltage difference must be reduced before connecting them. Excessive voltage difference can cause instantaneous currents of hundreds to thousands of amperes, damaging the system.
\end{enumerate}

\begin{figure}[htb]
  \centering
  \includegraphics[width=\columnwidth]{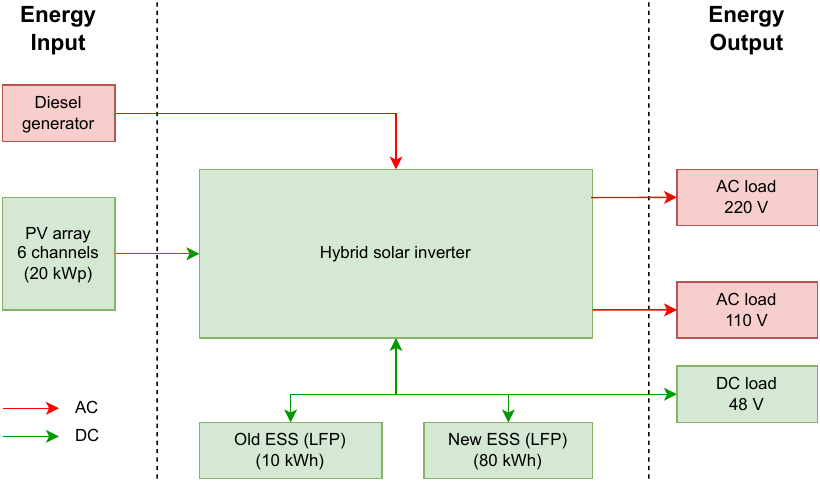}
  \caption{
  \justifying
  \textbf{Future energy architecture of Jiujiu Cabins (planned in 2021).}
  On the energy input side, PV arrays will be the primary energy source, while the generator will be an auxiliary system, supplementing the required power when solar energy supply is insufficient or malfunctions. On the energy output side, power will be provided for 220 V$_\mathrm{AC}$ loads, 110 V$_\mathrm{AC}$ loads, and 48 V$_\mathrm{DC}$ loads.
  }
  \label{fig:EnergyArchitecture-Future}
\end{figure}

After system adjustments, an off-grid solar ESS will be established to achieve the purposes of energy generation, storage, and consumption. Solar energy will be the primary energy source, while the diesel generator will serve as a backup. The LFP ESS will be used for energy storage, supplying energy to the cabins' loads via hybrid solar inverters. The system architecture is shown in Fig.~\ref{fig:EnergyArchitecture-Future}. On the left side, at the energy input, PV arrays will be the primary energy source, while the generator will be an auxiliary system, supplementing the required power when the solar energy supply is insufficient or malfunctions. On the right side, at the energy output, power will be provided for 220 V$_\mathrm{AC}$ loads, 110 V$_\mathrm{AC}$ loads, and 48 V$_\mathrm{DC}$ loads.

Regarding power output, the cabins originally used 110 V$_\mathrm{AC}$, but considering the potential use of high-power equipment in the future, 220 V$_\mathrm{AC}$ power will be added. For any potential additions of DC equipment, such as radio devices, 48 V$_\mathrm{DC}$ power will be provided, reducing losses caused by AC/DC conversion. The application scenarios for each voltage are briefly described below: 110 V$_\mathrm{AC}$ is the most common standard voltage in Taiwanese homes and offices, suitable for most low-power household appliances and electronic devices. Main uses include lighting fixtures, televisions, computers, mobile phone charging, electric fans, vacuum cleaners, small refrigerators, microwave ovens, and other everyday appliances. 220 V$_\mathrm{AC}$ is mainly used in Taiwan for high-power appliances to improve efficiency and reduce current load, ensuring electrical safety. Common uses include air conditioners, water heaters, induction cookers, large ovens, dryers, and some imported appliances. 220 V$_\mathrm{AC}$ halves the current, reducing line losses. 48 V$_\mathrm{DC}$ is mainly used in telecommunications infrastructure, light electric vehicles, industrial automation, and high-efficiency solar energy storage systems, offering advantages such as high power density, low transmission loss, and high safety. Common applications include data center routers, automotive 48 V$_\mathrm{DC}$ mild hybrid systems, warehouse robots, and automated robotic arms.

Junction boxes are divided into three main categories: PV panel, AC panel, and DC panel. The PV panel provides DC power transmission between the PV arrays and hybrid solar inverters. The AC panel provides power transmission from the hybrid solar inverters to 220 V$_\mathrm{AC}$ loads, from the generator to the hybrid solar inverters, and from the 220 V$_\mathrm{AC}$ power to the AC transformer, converting the 220 V$_\mathrm{AC}$ power to supply 110 V$_\mathrm{AC}$ loads. The DC panel provides bidirectional power transmission between the hybrid solar inverters and the LFP ESS.

Additionally, a self-developed cloud-native EMS can detect PV power generation (unit: W), PV energy generation (unit: kWh), load AC power consumption (unit: W), and load AC energy consumption (unit: kWh), showing the values on the enclosure using meters with liquid crystal display (LCD). The information is also uploaded to a cloud server for remote monitoring.

\section{Summary and Outlook}
\label{sec:SummaryOutlook}

Using renewable energy and eco-friendly ESS in high mountains is not easy.
It took about two decades to approach the goal in Jiujiu Cabins.
Since 2003, renewable energy provided by PV systems has been used.
However, the lead-acid battery ESS still causes pollution in high mountains.
Until 2021, encountering a system failure and repairing period, the eco-friendly LFP ESS replaces part of the lead-acid battery ESS.

In the next step, the remaining lead-acid batteries should be replaced by the eco-friendly LFP ESS entirely, making a better and greener off-grid solar ESS.
On the other hand, a more comprehensive EMS should be installed to check the energy input and output.
Figuring out the energy usage of the cabins.
Further suggestions can be given after the energy usage analysis, such as ``Is the energy generated sufficient? Is it necessary to expand the PV system?'' or ``Is the energy stored sufficient? Is it necessary to expand the ESS?''

Since the Executive Yuan's ``Open forest policy'' was implemented in Taiwan in 2019, renovations and equipment upgrades have begun on more than 40 mountain huts. Regarding renewable energy use, solar power is the primary source, with 90\% of the mountain huts already equipped with it. However, ESSs still use lead-acid batteries, which are not only highly polluting but also have short lifespans, making them prone to system downtime. Switching to LFP batteries for ESSs would not only be environmentally friendly but also extend the overall lifespan of the system.

We wish not only for a green building in high mountains, but also for using green energy to protect the ecosystem and environment.

\appendices
\section{\break Several notes related to the software integration of the cloud system}

In terms of software and system design, there are several points to note:
\begin{itemize}
  \item Integrate cloud-native energy information for over 40 energy nodes.
  \item Design an energy information dashboard for the management station. Design an energy information dashboard for engineers.
  \item Increase database space to handle the increased data volume.
  \item Increase the number of cloud server snapshots, maintaining at least two snapshots to ensure data maintenance.
  \item Add a second set of cloud servers and databases to maintain the stability of cloud data delivery.
\end{itemize}

Regarding cloud environments, there are several points to note:
\begin{itemize}
  \item Computing units should be hosted on a reliable and scalable infrastructure. A Service-Level Agreement (SLA) commitment with 99.99\% availability is required, with capacity scalable within minutes. A secure computing environment should also be provided.
  \item High-performance block storage is an easy-to-use, scalable, designed for use with computing units. It features rapid scaling to meet the most demanding high-performance workloads, including mission-critical applications such as SAP, Oracle, and Microsoft products. It prevents failures with high availability, including write-through within available zones (AZs) and 99.999\% durability using io2 Block Express disks. The disks offer high performance with high IOPS and throughput.
  \item Virtual cloud environments allow administrators to define and launch cloud resources within logically isolated virtual networks. Administrators have complete control over the virtual network environment, including resource placement, connectivity, and security. This is achieved by configuring the virtual cloud environment in the service console. Next, computing units and high-performance block storage are configured. Finally, the communication between virtual cloud environments, across accounts, and within availability zones is defined for zone communication.
  \item The load balancer protects applications using SSL/TLS termination, integrated credential management, and client-side credential verification. It provides highly available and automatically scaling applications. It monitors application operational status and performance in real time, identifies bottlenecks, and maintains SLA compliance.
  \item The data storage service used should be scalable, data available, secure, and efficient. It should protect any amount of data, such as data lakes, websites, cloud-native applications, backups, archives, machine learning, and analytics. Its storage service should be designed to achieve 99.999999999\% (11 nines) durability.
\end{itemize}

\section*{Acknowledgment}
The author (H.-C. Chung) would like to thank the contributors to this article for their valuable discussions and recommendations: Jung-Feng Jack Lin, Hsiao-Wen Yang, Yen-Kai Lo, and An-De Andrew Chung.
The author (H.-C. Chung) thanks Pei-Ju Chien for English discussions and corrections, as well as Ming-Hui Chung, Su-Ming Chen, Lien-Kuei Chien, and Mi-Lee Kao for financial support.
This work was supported in part by Super Double Power Technology Co., Ltd., Taiwan, under the project ``Development of Cloud-native Energy Management Systems for Medium-scale Energy Storage Systems (\href{https://osf.io/7fr9z/}{https://osf.io/7fr9z/})'' (Grant number: SDP-RD-PROJ-001-2020).


%
%

\begin{CJK}{UTF8}{bsmi}


\end{CJK}

\begin{IEEEbiography}
[{\includegraphics[width=1in,height=1.25in,clip,keepaspectratio]{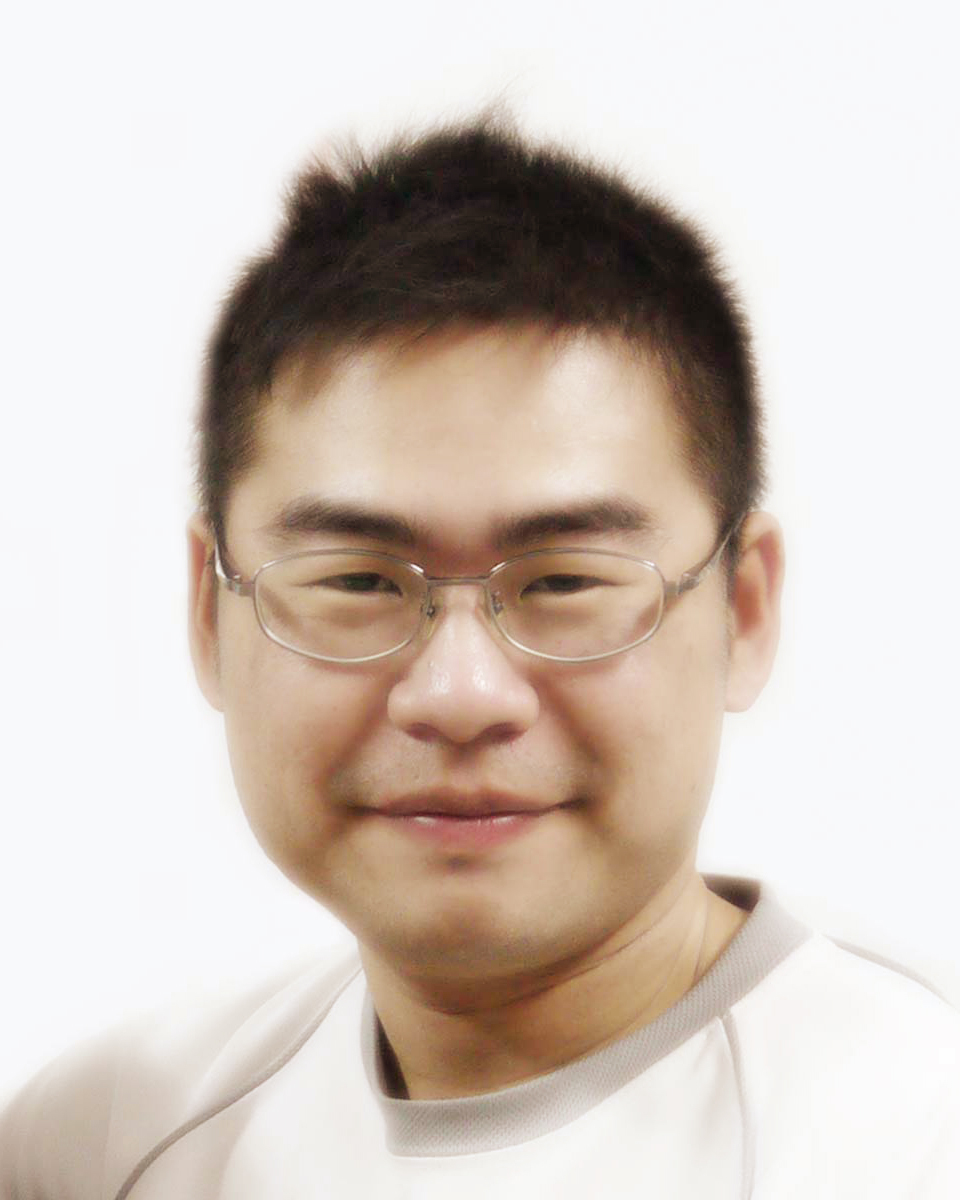}}]
{Hsien-Ching Chung} (M'17--SM'22) received the B.S., M.S., and Ph.D.  degrees in physics from the National Cheng Kung University, Tainan, Taiwan, in 2011.

Dr. Chung has complete experience in the lithium-ion (Li-ion) battery industry supply chain, from the battery cell factory, pack factory, to the system integration factory. Recent interests focus on system integration and applications of Li-ion battery energy storage systems. Currently, he is conducting the RD project "Development of Cloud-native Energy Management Systems (EMSs) for Medium-scale Energy Storage Systems." The cloud-native EMS has been installed at the Paiyun lodge (the highest lodge) in Taiwan.

He also has rich experience in fundamental research. From 2011 to 2017, as a postdoctoral fellow, his main scientific interests in condensed matter physics included the electronic and optical properties of carbon-related materials, low-dimensional systems, and next-generation energy materials.

Dr. Chung is a senior member of the Institute of Electrical and Electronics Engineers (IEEE), a member of the American Physical Society (APS), a member of the American Chemical Society (ACS), and an associate member of the Institute of Physics (IOP).
\end{IEEEbiography}

\EOD

\end{document}